# Comment on "Quantitative Model for the Superconductivity Suppression in $R_{1-x}Pr_xBa_2Cu_3O_7$ with Different Rare Earths"

In a recent letter [1], a model for the electronic structure of $R_{1-x}Pr_xBa_2Cu_3O_7$, based on the LDA + U scheme was proposed. The $T_c$ suppression by Pr substitution was argued to be due to the transfer of holes from the superconducting CuO $pd\sigma$ to a *dispersive $pf\pi$* band formed by O $2p_\pi$ and Pr $4f$ orbitals. The insulating nature of $PrBa_2Cu_3O_7$ was ascribed to disorder from Ba atoms on the Pr sites. In contrast to Ref. [2], where a similar model starting from a local description was proposed, the calculated width $W$ of the $pf\pi$ LDA band is substantial ($\sim 1.6$eV). Furthermore, it was argued that Ref. [2] fails to explain the ion size effect [3] on the critical concentration $x_c$ at which $T_c$ vanishes for different R in $R_{1-x}Pr_xBa_2Cu_3O_7$.

We comment on three points: (i) The real part of the ac conductivity $\sigma_1(\omega)$ from a single crystal of $PrBa_2Cu_3O_7$ [4] exhibits a large anisotropy. The spectral weight with polarization parallel to the chains is substantially larger than that perpendicular to the chains. The latter probes the response of the planar states. If those states were well described by a single-particle model like the $pf\pi$ LDA band mentioned above, disorder on the R site (claimed to be responsible for the localization of these otherwise highly dispersive states) should lead to a Drude-like ac conductivity for frequencies $\omega$ larger than a small cut-off frequency $\omega_c \ll W$. For $\omega < \omega_c$, the conductivity should drop to zero, due to the localization of single-particle states in 2D.

Fig. 1 shows $\sigma_1(\omega)$ for a nearest neighbour tight-binding model ($t = 0.2$eV, *i.e.*, $W = 1.6$eV) on an $L \times L$ ($L = 50$) square lattice in the presence of potential disorder, modelled by changing the on-site energies of N randomly selected sites to some value $V$. The conductivity was obtained from

$$\sigma_1(\omega) = \frac{\pi e^2}{L^2} \sum_{m \neq 0} \frac{|\langle \Psi_m | j_x | \Psi_0 \rangle|^2}{E_m - E_0} \delta[\omega - (E_m - E_0)], \quad (1)$$

where $j_x$ the current operator, and $E_0, E_m$ ($\Psi_0, \Psi_m$) the exact eigenvalues (eigenstates) from numerical diagonalization. The impurity concentration was $n_i = N/L^2 = 0.1$, the electron density $n_e = 0.6$, corresponding to a complete transfer of doped holes from the two (one for each $CuO_2$ sheet) $pd\sigma$ bands to the single $pf\pi$ band. A data point at frequency $\omega$ was obtained by adding the weight of all excitations within an interval $[\omega - \delta\omega, \omega + \delta\omega]$ ($\delta\omega = 0.006t$) and averaging over 10 realizations of the disorder. The plots clearly show Drude-like behaviour (with a *large* low-frequency conductivity) as argued above, in contradiction to the experimental spectra [4], which exhibit a slowly rising *small* conductivity up to energies of the order of 3eV. Hence, we conclude that a single-particle model with potential disorder as proposed in [1] *cannot explain* the ac conductivity in $PrBa_2Cu_3O_7$, and argue that strong correlations (best described by a *local* picture as outlined in [2]) are primarily responsible for the absence of Drude conductivity.

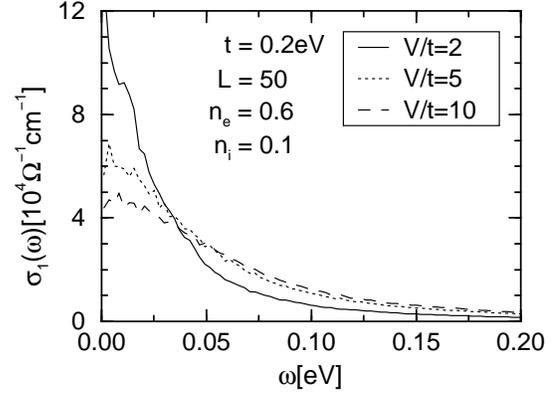

**Fig. 1.** Ac conductivity of a 2D tight-binding model on a square lattice in the presence of potential disorder.

(ii) In Ref. [2], the absence of superconductivity in $PrBa_2Cu_3O_7$ is due to the existence of a $Pr^{IV}$ oxidation state formed from hybridized Pr $4f$ and O $2p_\pi$ orbitals which lowers the energy of added holes as compared to the superconducting CuO $pd\sigma$ band. The energy difference $\Delta_{PF}$ between these two states is dependent on essentially two interatomic distances: $d_{CuO}$, the planar CuO distance, and $d_{RO}$, the distance between the rare earth atom and its O neighbours. An analysis of these distances for the series $RBa_2Cu_3O_7$ [5] shows that $\Delta_{PF}$ *decreases with decreasing ionic size of R*, favouring the superconducting solution. Hence, the model of Ref. [2] is in *qualitative agreement* with the observed increase of the critical concentration $x_c$ with decreasing R ion size in $R_{1-x}Pr_xBa_2Cu_3O_7$ [3].

(iii) The splitting between the spin-minority and spin-majority channels of the $pf\pi$ LDA band in [1] is of the order of its width $W$. This is usually an indication for the importance of correlation effects, and underlines our conclusions in (i).

We acknowledge financial support by the NSF (NSF-DMR-91-20000) through the Science and Technology Center for Superconductivity.

R. Fehrenbacher[1] and T. M. Rice[2,3]

[1] Materials Science Division
Argonne National Laboratory
Argonne, Illinois 60439

[2] AT&T Bell Laboratories
Murray Hill, New Jersey 07974

[3] ETH Hönggerberg
CH-8093 Zürich, Switzerland

[1] A. I. Liechtenstein and I. I. Mazin, Phys. Rev. Lett. **74**, 1000 (1995).
[2] R. Fehrenbacher and T. M. Rice, Phys. Rev. Lett. **70**, 3471 (1993).
[3] Y. Xu and W. Guan, Phys. Rev. B **45**, 3176 (1992).
[4] K. Takenaka et al., Phys. Rev. B **46**, 5833 (1992).
[5] M. Guillaume et al., Z. Phys. B **90**, 13 (1993).